\documentclass[aps,prl,preprint,tightenlines,superscriptaddress,showpacs,byrevtex]{revtex4}

\usepackage{graphicx}
\usepackage{dcolumn}
\usepackage{bm}

\graphicspath{{figs/}}

\def\jpsi {J/\psi}
\def\lmd  {\Lambda}

\def\b+ {B^{+}}
\def\b- {B^{-}}
\def\pbar{\bar{p}}

\def\DE {\Delta E}

\def\PB {\vec{p}_{B}}
\def\row {\rightarrow}
\def\DMB {\Delta M_{B}}
\def\Mbc {M_{\rm bc}}    

\begin{document}

\begin{minipage}[t]{4cm}
\includegraphics[width=3cm]{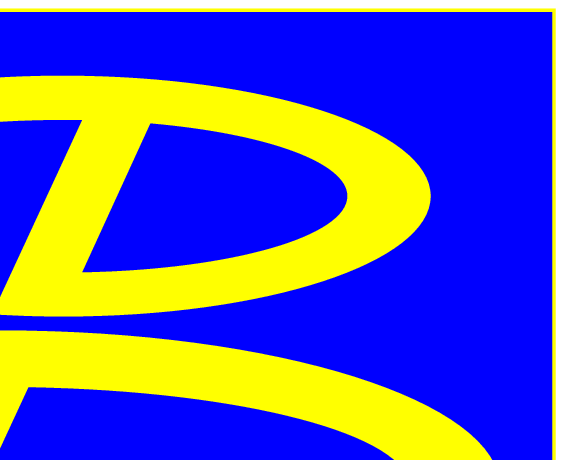}
\end{minipage}
\begin{minipage}[t]{10cm}
\vskip -2.2cm
\hspace*{6cm}{KEK Preprint 2003-60}\\
\hspace*{6cm}{Belle Preprint 2003-18}
\end{minipage}


\vspace{ 1cm }			  
\title{\boldmath
Search for $\b- \row \jpsi \lmd \pbar$ decay
}

%
%
\date{\today} 
%
\affiliation{Budker Institute of Nuclear Physics, Novosibirsk}
\affiliation{Chiba University, Chiba}
\affiliation{University of Cincinnati, Cincinnati, Ohio 45221}
\affiliation{University of Frankfurt, Frankfurt}
\affiliation{Gyeongsang National University, Chinju}
\affiliation{University of Hawaii, Honolulu, Hawaii 96822}
\affiliation{High Energy Accelerator Research Organization (KEK), Tsukuba}
\affiliation{Hiroshima Institute of Technology, Hiroshima}
\affiliation{Institute of High Energy Physics, Chinese Academy of Sciences, Beijing}
\affiliation{Institute of High Energy Physics, Vienna}
\affiliation{Institute for Theoretical and Experimental Physics, Moscow}
\affiliation{J. Stefan Institute, Ljubljana}
\affiliation{Kanagawa University, Yokohama}
\affiliation{Korea University, Seoul}
\affiliation{Kyungpook National University, Taegu}
\affiliation{Institut de Physique des Hautes \'Energies, Universit\'e de Lausanne, Lausanne}
\affiliation{University of Ljubljana, Ljubljana}
\affiliation{University of Maribor, Maribor}
\affiliation{University of Melbourne, Victoria}
\affiliation{Nagoya University, Nagoya}
\affiliation{Nara Women's University, Nara}
\affiliation{National Kaohsiung Normal University, Kaohsiung}
\affiliation{National Lien-Ho Institute of Technology, Miao Li}
\affiliation{Department of Physics, National Taiwan University, Taipei}
\affiliation{H. Niewodniczanski Institute of Nuclear Physics, Krakow}
\affiliation{Nihon Dental College, Niigata}
\affiliation{Niigata University, Niigata}
\affiliation{Osaka City University, Osaka}
\affiliation{Osaka University, Osaka}
\affiliation{Panjab University, Chandigarh}
\affiliation{Peking University, Beijing}
\affiliation{Princeton University, Princeton, New Jersey 08545}
\affiliation{RIKEN BNL Research Center, Upton, New York 11973}
\affiliation{Saga University, Saga}
\affiliation{University of Science and Technology of China, Hefei}
\affiliation{Seoul National University, Seoul}
\affiliation{Sungkyunkwan University, Suwon}
\affiliation{University of Sydney, Sydney NSW}
\affiliation{Tata Institute of Fundamental Research, Bombay}
\affiliation{Toho University, Funabashi}
\affiliation{Tohoku Gakuin University, Tagajo}
\affiliation{Tohoku University, Sendai}
\affiliation{Department of Physics, University of Tokyo, Tokyo}
\affiliation{Tokyo Institute of Technology, Tokyo}
\affiliation{Tokyo Metropolitan University, Tokyo}
\affiliation{Tokyo University of Agriculture and Technology, Tokyo}
\affiliation{Toyama National College of Maritime Technology, Toyama}
\affiliation{University of Tsukuba, Tsukuba}
\affiliation{Utkal University, Bhubaneswer}
\affiliation{Virginia Polytechnic Institute and State University, Blacksburg, Virginia 24061}
\affiliation{Yonsei University, Seoul}
\author{S.~L.~Zang}\affiliation{Institute of High Energy Physics, Chinese Academy of Sciences, Beijing} 
\author{K.~Abe}\affiliation{High Energy Accelerator Research Organization (KEK), Tsukuba} 
\author{K.~Abe}\affiliation{Tohoku Gakuin University, Tagajo} 
\author{T.~Abe}\affiliation{High Energy Accelerator Research Organization (KEK), Tsukuba} 
\author{I.~Adachi}\affiliation{High Energy Accelerator Research Organization (KEK), Tsukuba} 
\author{Byoung~Sup~Ahn}\affiliation{Korea University, Seoul} 
\author{H.~Aihara}\affiliation{Department of Physics, University of Tokyo, Tokyo} 
\author{M.~Akatsu}\affiliation{Nagoya University, Nagoya} 
\author{Y.~Asano}\affiliation{University of Tsukuba, Tsukuba} 
\author{T.~Aso}\affiliation{Toyama National College of Maritime Technology, Toyama} 
% ...
\collaboration{The Belle Collaboration}

\begin{abstract}
We report the results of a search for $B^{-} \row \jpsi \lmd \pbar$ based on 
a data set of 78 fb$^{-1}$ data collected at the $\Upsilon(4S)$ resonance 
with the Belle detector at the KEKB asymmetric $e^+ e^-$ collider.
No substantial signal is found,
and we set the branching fraction upper limit 
 ${\cal B}(B^{-} \row \jpsi \lmd \pbar)<4.1 \times 10^{-5}$  
at $90\%$ confidence level.
\end{abstract}


\maketitle

The inclusive decay of $B \row \jpsi + X$ has been 
studied by CLEO\cite{BA95}, Belle\cite{SS00}, and recently BaBar\cite{Au02}.
The $\jpsi$ momentum spectrum in the $e^{+}e^{-}$ center of mass 
frame  is consistent with the distribution predicted by non-relativistic 
QCD calculations\cite{BSW00}, except for an excess in the low momentum region 
which has been observed by all of the above experiments.
The excess below 0.8 GeV/$c$ corresponds to a branching fraction 
of $4\times 10^{-4} - 6\times 10^{-4}$.

In order to explain this excess, several theoretical hypotheses have been
proposed\cite{BN97,HC01,ELY02}. One of them is that the excess 
arises from $B \row \jpsi \lmd \pbar$ decays\cite{BN97}; 
this possibility can also be inferred from the rather large branching
fraction of $B \row \lmd \pbar X$, ($2.3 \pm 0.4$)\% \cite{PDG}.
At the quark level, $\b- \row \jpsi \lmd \pbar$ can be described as 
($ud\bar{u}\bar{d}$) produced by gluon emission from the Cabibbo favoured 
$b \row c \bar{c} s$ diagram. The decay rate could be enhanced by an 
intermediate exotic state such as a bound state of $\lmd$ and $\pbar$,  
$\jpsi$ and $\lmd$, or $\jpsi$ and $\pbar$.  In this case, the momentum 
distributions of the daughter particles will exhibit some characteristic 
enhancements.  Thus, searching for $\b- \row \jpsi \lmd \pbar$  helps 
to understand the source of the excess at low $\jpsi$ momentum and  
to find intermediate states.  The BaBar collaboration has 
recently reported results of a similar search\cite{Ba03}.

In this paper we report on the study of $\b- \row \jpsi \lmd \pbar$.
The analysis is based on a data sample of 78 fb$^{-1}$ accumulated 
at the $\Upsilon(4S)$ resonance with the Belle detector at the KEKB 
8 GeV $e^{-}$ and 3.5 GeV $e^{+}$ asymmetric collider\cite{KK03}.

The Belle detector consists of a three-layer silicon vertex detector (SVD), 
a 50-layer central drift chamber (CDC), an array of aerogel threshold 
$\rm\check{C}erenkov$ counters (ACC), time-of-flight scintillation 
counters (TOF), a CsI(Tl) crystal electromagnetic calorimeter (ECL), 
a 1.5 T superconducting solenoid coil and an instrumented iron-flux return 
for muon and $K_L$ detection (KLM).  The detector is described in detail
elsewhere\cite{Belle}.

In this analysis, we use the decay chain: $\b- \rightarrow J/\psi \lmd \pbar$, 
$J/\psi \rightarrow l^{+} l^{-}$($l$=$e$,$\mu$), and $\lmd \rightarrow p \pi^{-}$. 
Inclusion of charge conjugate states is implied throughout this paper. 
To suppress continuum backgrounds
we require the ratio of the second to zeroth Fox-Wolfram
moments\cite{FoxWolf} to be less than 0.5.

To remove charged tracks that are badly measured or do not come from the 
interaction region, we require leptons from $J/\psi$ to originate from 
within 5 cm of the interaction point along the beam direction. 
Both lepton tracks are required to be well identified as leptons.
Electrons are identified using a combination of specific ionization 
measurements ($dE/dx$) from the CDC,  the ACC response, and
electromagnetic shower position, shape and energy from the ECL\cite{HK02}. 
Muons are identified with KLM hit positions and penetration depth\cite{AA02}. 
In order to recover dielectron events where one or both electrons have radiated a
photon (final state radiation or bremsstrahlung), we include the four-momentum 
of every photon detected within 0.05 radians of the original $e^+$ or $e^-$ 
direction in the invariant mass calculation. The invariant mass of the candidate
$J/\psi\rightarrow\mu^+\mu^-(e^+e^-)$ is required to be between 
$-60(-150)$ MeV/$c^2$ and $+36(+36)$ MeV/$c^2$ of the known $J/\psi$ mass. 
The asymmetric mass requirements are due to radiative tails. 
The selection criteria for $J/\psi$ are identical to those used in ref.\cite{ABE03}.

Particle identification information from the ACC, TOF and $dE/dx$ information from the CDC 
is used to construct likelihoods $L_i$ for each hadron 
type $i$ ($i$ = $\pi$, $K$, and $p$). We require
$L_p/(L_p+L_{K})> 0.6$ and $L_{\pi}/(L_{\pi}+L_{K})> 0.6$ 
to select protons and pions, respectively.
For the prompt $\pbar$ candidates, tracks that are positively identified 
as muons or electrons are rejected.

$\Lambda$ candidates are reconstructed via the $p\pi^-$ decay channel. 
We require that the transverse impact parameters of both $\lmd$ 
daughter tracks with respect to the nominal beam axis be greater than
0.03 cm, the $z$ distance between the daughter tracks before constraining 
the $\lmd$ vertex be less than 12 cm, and the vertex fitting $\chi^{2}$ 
be smaller than 100. The invariant mass of the $\Lambda$ candidate is 
required to be within $\pm$6 MeV/$c^2$ of the nominal $\Lambda$ mass.
These criteria are determined to maximize $S/\sqrt{S+B}$, where $S$ is
the number of expected signal events in the signal region defined below
obtained from Monte Carlo (MC), and $B$ is the number of expected
background events obtained from sideband data.  For the calculation of
$S$, we assume the branching fraction of $\b- \rightarrow J/\psi \lmd
\pbar$ to be $1.0 \times 10^{-5}$ which is consistent with our final result.
We apply vertex and mass constrained fits for $\jpsi$ and $\Lambda$
candidates to improve the momentum resolution.

We identify $\b- \rightarrow J/\psi \lmd \pbar$ candidates using
two kinematic variables calculated in the center of mass frame: the beam-energy 
constrained mass, $\Mbc \equiv \sqrt{E_{\rm beam}^{2}-\PB^{2}}$, and 
the mass difference, $\DMB \equiv M_{B}- m_{B^-}$, 
where $E_{\rm{beam}}$ is the beam energy, $\PB$ and 
$M_{B}$ are the reconstructed momentum and mass of the $B$ candidate, 
and $m_{B^-}$ is the nominal $B^-$ mass. 
Candidates with $|\DMB|<0.20$ GeV/$c^{2}$ and $\Mbc >5.20$ GeV/$c^{2}$ 
are selected for the final analysis. The signal region is defined to be 
5.27 GeV/$c^2 < \Mbc <$ 5.29 GeV/$c^2$ and $|\DMB| < 0.03$ GeV/$c^2$, 
which corresponds to three standard deviations based on the MC simulation.
We also define a $\DMB$ sideband region as 0.06 GeV/$c^2 < \Delta M_B <$ 0.20 GeV/$c^2$.

In this analysis, we do not use the widely used $\DE \equiv E_{B}-E_{\rm beam}$,
where $E_{B}$ is the energy of the reconstructed $B$.
For the $\jpsi \lmd \pbar$ events, the kinematic limit of $\DE$ can be expressed as 
$\DE = \sqrt{M_{\rm tot}^{2}+E_{\rm beam}^2-\Mbc^{2}}-E_{\rm beam}$, 
where $M_{\rm tot}$ is the sum of the masses of $\jpsi$, $\lmd$ and $\pbar$. 
This kinematic limit is close to the signal region and depends on $\Mbc$, which
introduces a distortion in the $\Mbc-\DE$ phase space. Consequently, events outside 
the $\Mbc/\DE$ signal region cannot be used to model the $\DE/\Mbc$ distributions 
for the background. Furthermore, we find a large negative correlation between 
$\DE$ and $\Mbc$ for signal events.  On the contrary, $\DMB$ and $\Mbc$ are 
uncorrelated, which is confirmed using Monte Carlo simulations.  
Therefore, we use $\DMB$ and $\Mbc$ to identify signal candidates 
in this analysis\cite{note1}. 

Around 15\% of the selected events have more than one $B$ candidate.  We select the
best candidate by first choosing the $\Lambda$ candidate with the smallest
total $\chi^2$ of the vertex and mass constrained fit.
If multiple $B$ candidates with the same $\Lambda$ remain, 
we select the one with the smallest $\chi^2$ of the $J/\psi \bar{p}$ vertex.

  \begin{figure*}[htp]
   \includegraphics[width=0.65\textwidth]{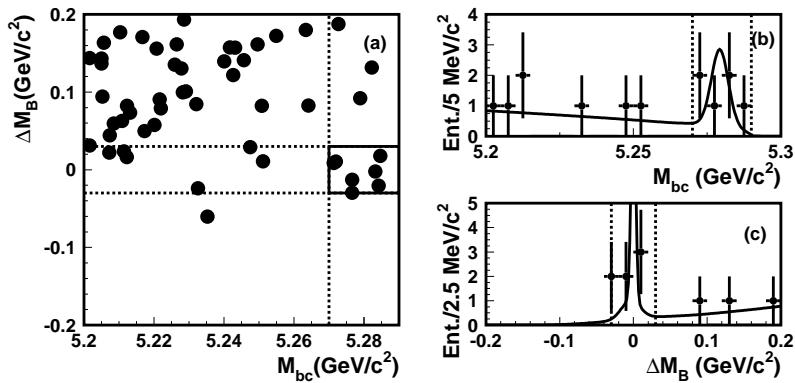}
   \caption{(a) ($\DMB,\Mbc$) distribution of $\b- \row \jpsi \lmd \pbar$ 
   candidates, and its projections onto (b) $\Mbc$ and (c) $\DMB$. The dashed 
   lines indicates the bands used for the projections. The curves illustrate the 
   result of the fit described in the text.}
    \label{scatter-plot}
   \end{figure*}

Figure~\ref{scatter-plot} shows a scatter plot of $\DMB$ versus $\Mbc$ and 
their projections for candidates after all selection criteria are applied. 
The $\DMB$($\Mbc$) projection is shown for candidates in the $\Mbc$($\DMB$) 
signal region. There are six events in the ($\Mbc$,$\DMB$) signal region.
We obtain the signal yield by fitting the $\Mbc$ distribution, since 
this variable has better agreement between data and MC than $\DMB$.  
The background is described by an ARGUS function\cite{ARG} and
the signal PDF is modelled by a sum of two Gaussians plus a Crystal Ball line 
function\cite{CB} to account for the small tail, with parameters obtained from MC.
In the fitting, $E_{\rm{beam}}$ and the width of the main Gaussian are fixed 
as 5.289 GeV and 2.62 MeV/$c^{2}$ respectively, based on a control sample 
of $B^- \to D^0 \pi^-$. 
We simultaneously fit the signal and sideband regions, where the PDF in
the sideband region is given by an ARGUS function with the same parameters
as used in the signal region, except for the normalization, which is allowed 
to float. The $\Mbc$ fit gives $3.5\pm2.3$ signal events and $0.92\pm0.34$ 
background events in the signal box. The statistical significance of 
the signal, defined as $\sqrt{-2\ln(L_{0}/L_{\rm max})}$, is 2.3,  
where $L_{\rm max}$ and $L_{0}$ denote the maximum likelihood with the
fitted signal yield and with the yield fixed at zero, respectively. 
We examine the contribution of other decay modes including $\jpsi$ and
baryons, such as $B^- \to \jpsi \Sigma^0 \pbar$, which may also peak in the
$\Mbc$ signal region. Using MC, we find them to be negligible with an assumption
that their branching fractions are comparable to $ B^- \to \jpsi \lmd \pbar$.
As a cross check we also fit the $\DMB$ distributions; the obtained signal yield 
is $4.7 \pm 2.7$, with a statistical significance of 2.8, while the background 
yield is $0.55 \pm 0.31$. These results are consistent with the results of 
the $\Mbc$ fit. Since the signal yield is not substantial,
we give an upper limit of the branching fraction as the main result of 
this analysis.

The reconstruction efficiency ($\epsilon$) is estimated using signal MC 
where a three-body phase-space model is employed. 
We obtain $\epsilon = (6.3_{-2.3}^{+2.7})\% $.
The sources and amounts of systematic 
uncertainties are summarized in Table~\ref{table-eff-sys-err}.
The uncertainty of the tracking efficiency is estimated by adding the momentum 
dependent single track systematic error. The uncertainty is $\sim 1\%$ for
leptons from $\jpsi$, $\sim 3.5\%$ and $\sim 1.6\%$ for low momentum pions and 
protons from $\lmd$, and $\sim 1.4\%$ for prompt protons from $B$.
The uncertainty of the $\Lambda$ reconstruction is determined by comparing 
the proper time distributions for data and MC simulation.
For the uncertainty due to modeling three-body decays in the phase space, 
we conservatively assign the maximum variation of the efficiencies among the 
slices of $M(J/\psi,\Lambda)$, $M(J/\psi,\bar{p})$, and $M(\Lambda,\bar{p})$.

\begin {table}[htp]
\begin {center}
\begin {tabular}{|c|c|}
\hline
Source  & Relative systematic error \\
\hline
Tracking     & $\pm 8.5\%$    \\
PID(proton and pion) & $\pm 8\%$ (3\% per $p$, 2\% per $\pi$) \\
Lepton identification & $\pm 4\%$ (2\% per lepton)    \\
$\lmd$ reconstruction & $\pm 6\%$ \\
Simulation modeling & $+41.3\%$ / $-34.1\% $ \\
MC statistics    & $\pm 1.8\% $          \\
\hline
Total  &   $+43.6\%$ / $-36.8\%$  \\
\hline
\end {tabular}
\caption {Summary of uncertainties in the reconstruction efficiency. }
\label{table-eff-sys-err}
\end {center}
\end {table}

We use the method of Feldman and Cousins\cite{FC98}, including the
uncertainties of the background and efficiency estimations\cite{JC03}, to
obtain a 90\% confidence interval for the branching fraction given by
$N_{S}/(\epsilon \times N_{B\bar{B}} \times {\cal B}(J/\psi \row l^{+}l^{-})
\times {\cal B}(\lmd \row p \pi^{-}))$,  
where $N_{S}$ is the signal yield, 
$N_{B\bar{B}}$ is the number of $B\bar{B}$ pairs, $(85.0 \pm 0.5)\times 10^{6}$, 
and the decay branching fractions ${\cal B}(J/\psi \row l^{+}l^{-})$ and 
${\cal B}(\lmd \row p \pi^{-})$ are taken from the world averages\cite{PDG}.
The fractions of neutral and charged $B$ mesons produced in $\Upsilon(4S)$ decays
are assumed to be equal.  With six observed candidates,  
$0.92 \pm 0.34$ background events, and the uncertainties
in $\epsilon$, $N_{B\bar{B}}$, and secondary branching fractions mentioned above,
we obtain an upper bound of the interval from this procedure of 
${\cal B}(\b- \row \jpsi \lmd \pbar)<4.1\times10^{-5}$, which
we interpret as a conservative estimate of the 90\% confidence upper limit
of the branching fraction\cite{note2}.
%

  \begin{figure*}[htp]
    \includegraphics[width=0.7\textwidth]{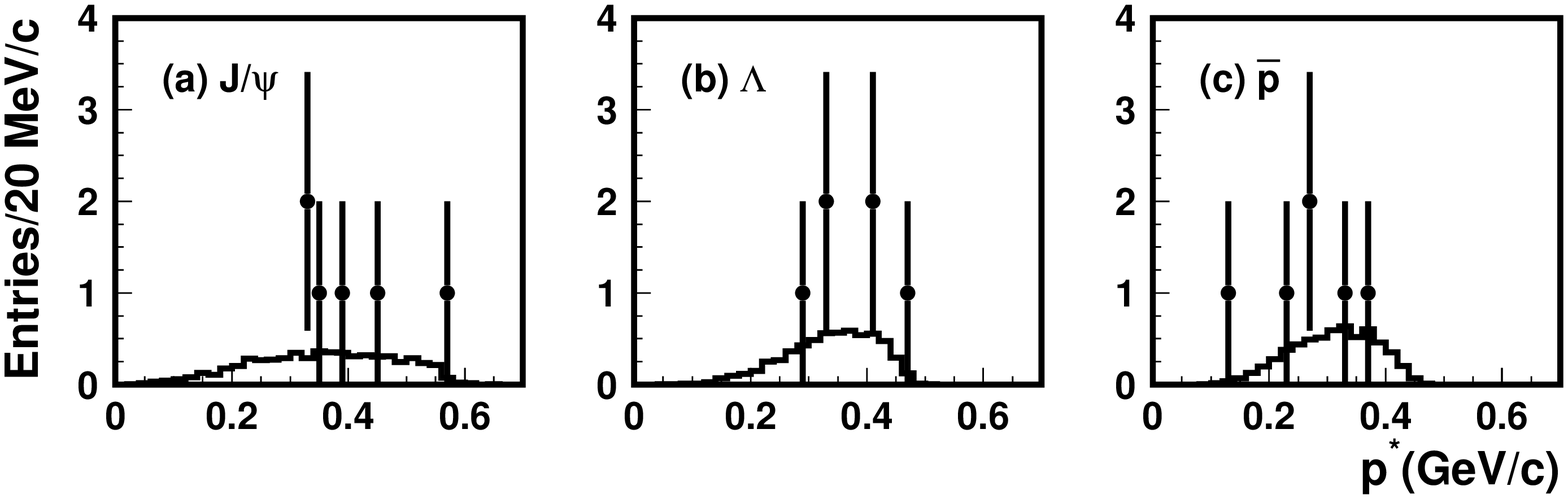}
   \caption{Momentum distributions of (a) $\jpsi$,
   (b) $\lmd$ and (c) $\pbar$ in the rest frame of the reconstructed $B$
    for the six $\b- \row \jpsi \lmd \pbar$ candidates. The histograms
    are phase space distributions from signal MC normalized to six events.}
     \label{momentum-plot}
  \end{figure*}
			
Figure~\ref{momentum-plot} shows the momentum distributions for $\jpsi$,
$\lmd$, and prompt $\pbar$ of the six candidates in the $B$ rest frame.
We do not observe any significant enhancement above the phase space 
distribution, such as would be expected by an intermediate resonance\cite{BN97}.

In summary, we have searched for the decay of $\b- \row \jpsi \lmd \pbar$ at 
Belle with 78 fb$^{-1}$ data collected at the $\Upsilon(4S)$ resonance. 
No statistically significant signals are found.  We set an upper limit of 
${\cal B}(\b- \row \jpsi \lmd \pbar)<4.1 \times 10^{-5}$ at $90\%$ confidence level. 
This result is consistent with the BaBar result\cite{Ba03}. This mode does not account 
for a significant fraction of the observed excess in the low momentum region 
of $B \to J/\psi X$.

  \begin{acknowledgments}
We wish to thank the KEKB accelerator group for the excellent
operation of the KEKB accelerator.
We acknowledge support from the Ministry of Education,
Culture, Sports, Science, and Technology of Japan
and the Japan Society for the Promotion of Science;
the Australian Research Council
and the Australian Department of Education, Science and Training;
the National Science Foundation of China under contract No.~10175071;
the Department of Science and Technology of India;
the BK21 program of the Ministry of Education of Korea
and the CHEP SRC program of the Korea Science and Engineering Foundation;
the Polish State Committee for Scientific Research
under contract No.~2P03B 01324;
the Ministry of Science and Technology of the Russian Federation;
the Ministry of Education, Science and Sport of the Republic of Slovenia;
the National Science Council and the Ministry of Education of Taiwan;
and the U.S.\ Department of Energy.
\end{acknowledgments}


\begin{thebibliography}{99}
\bibitem{note1}
    Further study shows this holds also for general $B$ 
    decay modes, and ($\DMB,M_{\rm bc}$) will be somewhat better than 
    ($\DE,M_{\rm bc}$) to isolate $B$ candidates.
\bibitem{note2}
   The 90\% confidence interval obtained is $[3.5\times10^{-6}, 41\times10^{-6}]$.
   The central value of the branching fraction is $(8.7^{+6.9}_{-6.5})\times10^{-6}$, 
   where the uncertainty includes both statistical and systematic components.
\end{thebibliography}
\end{document}